\documentclass[aps,prc,reprint,showpacs,showkeys,unsortedaddress]{revtex4-1}

\usepackage{graphicx}
\usepackage{dcolumn}
\usepackage{bm}
\usepackage{amsmath}
\usepackage{float}

\begin{document}

\title{Importance of realistic phase space representations of initial quantum fluctuations using the stochastic mean-field approach for fermions}

\author{Bulent Yilmaz}
\email[]{bulent.yilmaz@science.ankara.edu.tr}
\affiliation{Physics Department, Faculty of Sciences, Ankara University, 06100, Ankara, Turkey}
\author{Denis Lacroix}
\affiliation{Institut de Physique Nucl\'eaire, IN2P3-CNRS, Universit\'e Paris-Sud, F-91406 Orsay Cedex, France}
\author{Resul Curebal}
\affiliation{Physics Department, Faculty of Sciences, Ankara University, 06100, Ankara, Turkey}

\date{\today}

\begin{abstract}
In the stochastic mean-field (SMF) approach, an ensemble of initial values for a selected set of one-body observables is formed by stochastic sampling from a phase-space distribution that reproduces the initial quantum fluctuations. Independent mean-field evolutions are performed with each set of initial values followed by averaging over the resulting ensemble. This approach has been recently shown to be rather versatile and accurate in describing the correlated dynamics beyond the independent particle picture. In the original formulation of SMF, it was proposed to use a Gaussian assumption for the phase-space distribution. This assumption turns out to be rather effective when the dynamics of an initially uncorrelated state is considered, which was the case in all applications of this approach up to now. Using the Lipkin-Meshkov-Glick (LMG) model, we show that such an assumption might not be adequate if the quantum system under interest is initially correlated and presents configuration mixing between several Slater determinants. In this case, a more realistic description of the initial phase-space is necessary. We show that the SMF approach can be advantageously combined with standard methods to describe phase-space in quantum mechanics. As an illustration, the Husimi distribution function is used here to obtain a realistic representation of the phase-space of a quantum many-body system. This method greatly improves the description of initially correlated fermionic many-body states. In the LMG model, while the Gaussian approximation failed to describe these systems in all interaction strength range, the novel approach gives a perfect agreement with the exact evolution in the weak coupling regime and significantly improves the description of correlated systems in the strong coupling regime.
\end{abstract}



\maketitle

\section{Introduction}
In recent years, the interest in describing slow and fast dynamics of many-body fermionic systems has strongly increased. This renewal of interest is largely due to the advances in experimental studies that allow to form and manipulate a limited set of fermions in the laboratory \cite{Blo08,Hac12,Sta12,Sto10}. Dedicated studies have been recently made on the fast dynamics of systems after a quantum quench \cite{Ber14, Sch12}. Most, if not all, quantum transport theories dedicated to interacting fermions use the mean-field theory as a starting point. Deterministic approaches, based on density matrix or non-equilibrium Green function (NEGF) evolution, that aim at including the effect of correlations turn out to be rather involved numerically and have been applied with varying predictive powers \cite{Bon98, dahlen07, puigvonfriesen09,Her14}. In parallel, several stochastic methods where noise is added to the mean-field, have been proposed (for a recent review see \cite{Lac14a}). Among these approaches, one has recently attracted more interest due to its simplicity and its rather surprisingly good predictive power. This approach, called stochastic mean-field (SMF) was historically introduced in the nuclear physics context \cite{Ayi08}. In recent years, it was firstly validated in schematic model in Ref. \cite{Lac12}, extended in Ref. \cite{Lac13} to describe small superfluid systems and shown to be competitive with most recent NEGF theories applied to lattice systems in Ref. \cite{Lac14b}.     

\begin{figure}[htb]
\includegraphics[width=0.6\linewidth]{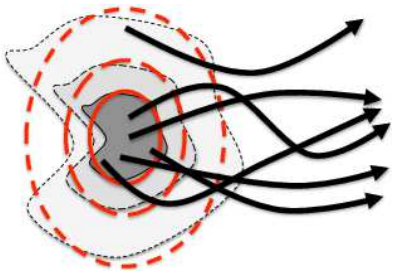}
\caption{(Color online) Schematic illustration of the SMF technique. Given a set of collective variables; sampling of their initial values from the initial phase-space associated to them (gray area) is made. Then, the sampled values are propagated with the associated mean-field equation of motion. Few examples of mean-field trajectories are presented with black thick lines. The red thick lines give a schematic picture of the Gaussian assumption made on the initial phase-space.}
\label{fig:schem}
\end{figure}

The SMF approach starts from the following observation. Time-dependent mean-field is generally adequate for describing the evolution of one-body collective observables but not their fluctuations especially in finite systems where zero point fluctuations can be large. In addition, even if fluctuations are accounted for at initial time, their effect on one-body evolution is poorly treated. As a matter of fact, TDHF can indeed be considered as a quasi-classical approximation for many-body fermionic systems where quantal fluctuations due to two-body correlations are poorly treated. The SMF approach is based on the fact that a quantum system can sometimes be accurately simulated by considering an ensemble of classical trajectories with initial values sampled from the initial phase-space \cite{Her84,Kay94}. The use of such a strategy, depicted in Fig. \ref{fig:schem}, is far from being straightforward in many-body systems. In the pioneering work of \cite{Ayi08}, the collective space for fermions was assumed to correspond to the full set of one-body observables. Then, four crucial assumptions have been made: (i) The initial phase-space has been obtained by imposing that the quantal average of first and second moments of one-body observables identifies with classical average. (ii) Consistently with the hypothesis (i), the sampling from the initial phase-space can be performed using a set of eventually correlated Gaussian random variables. (iii) The mean-field dynamics is used to propagate each initial conditions. (iv) At each time step, quantal fluctuations can be estimated using classical average. In Ref. \cite{Ayi08}, some formal arguments were given to support these ideas. Afterwards, it was shown in different test cases that the SMF concept \cite{Lac12,Lac13,Lac14a} can be more predictive and versatile than anticipated. In the present article, we analyze in more details the hypothesis made for the initial sampling, i.e. (i) and (ii). In particular, we show that such hypothesis seems to be rather effective if the initial state is uncorrelated, i.e. if it is a pure Slater determinant state. The accurate description of correlated systems requires to go beyond the Gaussian approximation for the initial phase-space. We show in that case, that SMF can be combined with the standard Husimi method to further increase its predictive power especially in the strong coupling case. Our discussion is illustrated with the Lipkin-Meshkov-Glick (LMG) model that has been used as the first sensitive test of SMF \cite{Lac12}.

\section{The Lipkin-Meshkov-Glick Model}
In this section, we will recall the minimal ingredients of the LMG model that are useful for the present discussion. More details can be found in Refs.
\cite{LMG,Rin80,Sev06,Zha90}.

\subsection{Exact evolution}
The LMG model was originally invented to represent the interaction of nucleons in two N-fold degenerate levels (one above and the other below the Fermi level) of a nucleus \cite{LMG}. It can also describe a system of N interacting spins in the xy plane under an influence of transverse magnetic field \cite{Botet,Vidal,Vidal2}. The LMG Hamiltonian is given by  
\begin{equation}
 H=\varepsilon J_z - V(J_x^2-J_y^2),
 \label{Ham}
\end{equation}
where $\varepsilon$ is the energy gap between the two states and $V$ is the interaction strength. We set $\hbar=1$. The quasispin operators are defined as 
\begin{gather}
J_x=\frac{1}{2}(J_+ + J_-),\quad J_y=\frac{1}{2i}(J_+ - J_-),\nonumber\\
J_z=\frac{1}{2}\sum_{n=1}^N \left( c^\dag_{+,n} c_{+,n}-c^\dag_{-,n} c_{-,n}\right),
\label{angmom}
\end{gather}
with
\begin{eqnarray}
 J_+=\sum_{n=1}^N c^\dag_{+,n} c_{-,n}\:,\quad J_-=J^\dag_+,
\end{eqnarray}
where $c^\dag_{+,n}$ and $c^\dag_{-,n}$ are creation operators of the upper and lower single-particle levels, respectively. The quasispin operators satisfy the usual SU(2) commutation relation, 
\begin{eqnarray}
 [J_x,J_y]=iJ_z,
\end{eqnarray}
and its cyclic permutations. The Hamiltonian in Eq. (\ref{Ham}) commutes with the total quasispin operator $J^2$ which has an eigenvalue $N/2(N/2+1)$. Hence the spin quantum number $j=N/2$ is conserved. Using the standard eigenvectors $|j\: m\rangle$ of $J^2$ and $J_z$ operators, exact solution of the LMG model can be obtained by forming the matrix representation of the Hamiltonian in the $|j\: m\rangle$ basis and diagonalizing the resulting $(N+1)\times(N+1)$ matrix.  

\subsection{The Mean-Field Approximation with SU(2) Spin Coherent States}
The mean-field approximation of the LMG model can be obtained by using the SU(2) spin coherent states \cite{Santos,Trimborn,GraefeKorsch,Kan80}. The spin (Bloch) coherent states are defined by
\begin{eqnarray}
 |\Omega\rangle=|\theta,\phi\rangle=\sum_{m=-j}^j&&\sqrt{2j \choose j-m}\left(\cos\frac{\theta}{2}\right)^{j+m}
 \left(\sin\frac{\theta}{2}\right)^{j-m} \nonumber \\
&&\times e^{i(j-m)\phi}\;|j\: m\rangle, 
\label{Bloch}
\end{eqnarray}   
where the two angles cover the parameter space $0\leq\theta\leq\pi$ and $0\leq\phi\leq 2\pi$ \cite{Gazeau,Vieira}. The coherent states are normalized, $\langle \Omega|\Omega\rangle=1$, and the closure relation is given by 
\begin{eqnarray}
 \frac{2j+1}{4\pi}\int |\Omega\rangle \langle\Omega| d\Omega=1, \label{eq:closure}
\end{eqnarray}
with $d\Omega=\sin\theta\,d\theta\,d\phi$. The coherent state algebra, discussed in detail in Refs. \cite{Zha90,Vieira}, can be used to get 
any moments of the spin operators. The expectation values of the first and second moments of quasispin operators are given by
\begin{eqnarray}
 \langle\Omega |J_x|\Omega\rangle&=&\frac{N}{2}\sin\theta\,\cos\phi,\nonumber\\
 \langle\Omega |J_y|\Omega\rangle&=&\frac{N}{2}\sin\theta\,\sin\phi,\nonumber\\
 \langle\Omega |J_z|\Omega\rangle&=&\frac{N}{2}\cos\theta,
 \label{firstmom}
\end{eqnarray}
and
\begin{eqnarray}
\langle \Omega |\widetilde{J_iJ_j}|\Omega\rangle&=&\frac{N-1}{N}\langle\Omega | J_i|\Omega\rangle\langle\Omega|J_j|\Omega\rangle+\frac{N}{4}\delta_{ij},
\label{secondmom}
\end{eqnarray}
where $\widetilde{AB}$ is the symmetrical ordering of operators, $\widetilde{AB}=(AB+BA)/2$ and $i,j=x,y,z$. By using the Ehrenfest theorem and Eqs. (\ref{firstmom}) and (\ref{secondmom}), the mean-field equations \cite{Kan80,Lac12}
\begin{eqnarray}
 \frac{d}{dt}j_x&=&\varepsilon\left(-j_y+\frac{2\chi}{N}j_y\,j_z\right),\nonumber\\
 \frac{d}{dt}j_y&=&\varepsilon\left(j_x+\frac{2\chi}{N}j_x\,j_z\right),\nonumber\\
 \frac{d}{dt}j_z&=&\varepsilon\left(-\frac{4\chi}{N}j_x\,j_y\right),
 \label{mf}
\end{eqnarray}
can easily be obtained where the expectation values given by Eq. (\ref{firstmom}) are replaced with 
c-number quasispins $j_i$. $\chi=V(N-1)/\varepsilon$ is a relative interaction strength parameter.  

The mean-field equations, Eq.  \eqref{mf}, can be derived either starting from a coherent state 
or from a statistical ensemble of coherent states. Then,   
the mean-field equations are solved with the initial values 
\begin{equation}
 j_i(0)={\rm Tr} \left(J_i\rho_0\right), 
 \label{mfini}
\end{equation}
where $\rho_0$ is the initial density matrix. If the initial state is a coherent state, we have $\rho_0=|\Omega(t=0)\rangle\langle\Omega(t=0)|$.   
As we will see below, the SMF approach allows to also make use of the mean-field equations, even if the initial state is correlated.  

\section{The Stochastic Mean-Field Approach}
In the mean-field approximation, time evolution of expectation values of the second and higher moments of collective observables depends only on the expectation values of the first moments, therefore quantum correlations of collective variables are mostly neglected leading to a classical picture. As demonstrated recently \cite{Lac14a}, the SMF approach can partially cure this deficiency. The assumption of SMF is that the dynamics of a many-body system can be approximated by an ensemble of independent mean-field evolutions starting from a statistical ensemble of initial values $j_i^\lambda(0)$ instead of the deterministic initial values given by Eq. \eqref{mfini}. Here, $\lambda$ labels a specific configuration of the ensemble. In the original formulation of SMF \cite{Ayi08} as well as in recent applications \cite{Lac12,Lac13}, it was proposed to sample the initial values using a Gaussian stochastic process in such a way that the first and second quantum moments of selected one-body observables, here $\{J_x,J_y,J_z\}$, are exactly reproduced at initial time. Then, each event in the resulting ensemble of initial values $\{j_x^\lambda(0),j_y^\lambda(0),j_z^\lambda(0)\}$ is evolved by the standard mean-field equations given by Eq. (\ref{mf}). The final results are obtained by averaging over the ensemble. This approach was shown to describe accurately the short-time evolution of many different quantum fermionic many-body systems \cite{Lac12,Lac13}.
 
Since the mapping between the quantal and classical problem is made by imposing solely the first two moments of a selected set of collective variables, the Gaussian assumption is a simple practical solution to the initial phase-space sampling problem and seems quite natural. However, in the following, we show that it might be sometimes advantageous to relax this hypothesis. Below, we first test the validity of the Gaussian assumption by considering alternative initial distributions.          

\subsection{Assumptions for the initial distribution of observables}
\begin{figure}[htb]
\includegraphics[width=7.5cm]{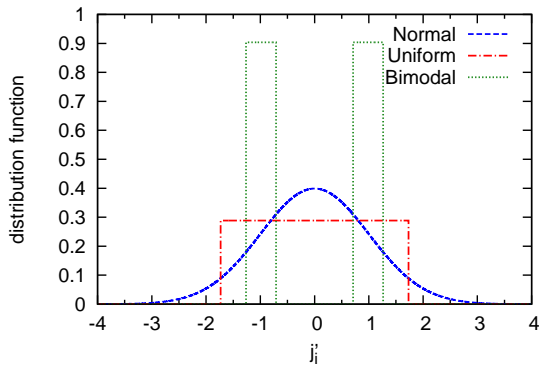}
\caption{(Color online) Illustration of the probability distributions used in the present study: normalized Gaussian (normal), uniform, and bimodal distributions in terms of the variable $j'_i=(j_i-\mu_i)/\sqrt{\sigma^2_{ii}}$. All distributions have zero mean and unit variance. 
The kurtosis for the bimodal distribution is arbitrarily chosen to be equal to 1.1.}
\label{fig:D} 
\end{figure}

Let us assume a simple distribution function, denoted generically by $D$, for the stochastic quantities $j_i^\lambda(0)$ that can reproduce the 
initial quantum expectation values of the first few moments of the observables. Hence we have
\begin{eqnarray}
\left[j_i^\lambda(0)\right]_D&=&\langle \psi_0|J_i|\psi_0\rangle,\nonumber \\
\left[j_i^\lambda(0)j_j^\lambda(0)\right]_D&=&\langle\psi_0|\widetilde{J_iJ_j}|\psi_0\rangle,\nonumber\\
&\vdots&
\label{vdots}
\end{eqnarray}
where $[...]_D$ stands for averaging over the assumed (predefined) distribution function. In general, it is possible to define infinitely many distributions that satisfy the expectation values of the same set of finite moments. Therefore, SMF approach is expected to depend on what distribution is chosen and how many expectation values are satisfied by the chosen distribution. For this purpose, we consider three distribution functions for the observables $j_i$, normal (Gaussian), uniform, and a bimodal distribution. Figure \ref{fig:D} shows the shapes of these distribution functions. The explicit forms and some properties of these distributions are given in Appendix A. The parameters of the all three distribution functions are arranged so that the same quantum means and (co)variances are reproduced. 

\subsection{Application}
In all the following computations, time is measured in units of $1/\varepsilon$. The number of events for stochastic sampling is $10^6$. The number of particles is $N=40$, hence $j=20$. For the sake of brevity, only one observable that best demonstrates the differences for compared conditions is plotted. The plots of the other observables, such as means and variances of quasispins, have similar differences for the compared conditions.  

\subsubsection{SMF evolution starting from a Slater determinant}
\begin{figure}[htb]
\includegraphics[width=7.5cm]{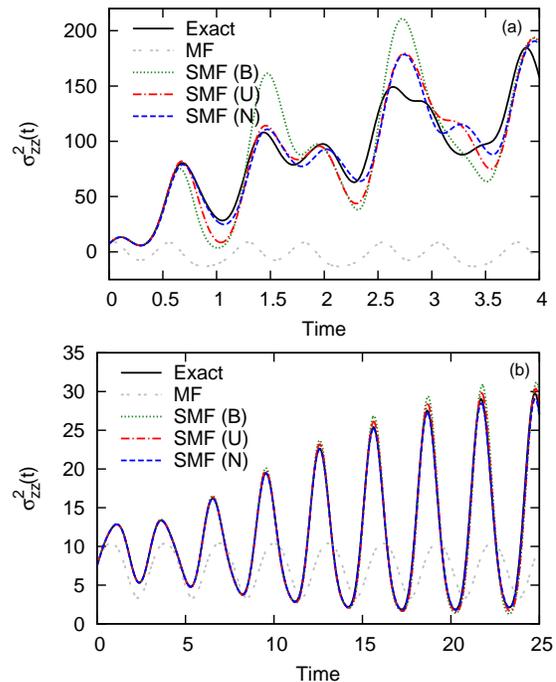}
\caption{(Color online) The variances of $J_z$ are plotted versus time for the initial coherent state $|\theta_0=\pi/3,\phi_0=\pi/4\rangle$. MF stands for mean-field and B, U, and N signify bimodal, uniform, and normal distributions, respectively. a) strong coupling ($\chi=5$) case. b) weak coupling ($\chi=0.5$) case.}
\label{fig:jz1}
\end{figure}

In Fig. \ref{fig:jz1}, the exact variance of the observable $J_z$ is compared with the variances computed with SMF approach using the three distributions. The initial state is the Bloch coherent state $|\psi_0\rangle=|\theta_0=\pi/3,\phi_0=\pi/4\rangle$ which is defined by Eq. (\ref{Bloch}). It is worth to mention that any Bloch coherent state corresponds to a Slater determinant in the LMG model \cite{Rin80}. Hence, Slater determinant and coherent state terminologies are interchangeably used in the text.    

Similarly to what has been observed in previous applications, the SMF approach applied to an initial coherent state turns out to be almost identical to the 
exact solution for weak two-body interactions. While a careful look to the SMF evolution shows that the use of Gaussian approximation for the initial distribution is slightly better than the use of uniform or bimodal distribution, we see that the evolution is overall insensitive to the assumed distribution in the weak-coupling regime. This is quite surprising in view of the completely different shapes of the distributions (see Fig. \ref{fig:D}). In contrary, in the strong coupling case, the evolution is quite sensitive to the assumption for the initial distribution. In that case, a much better agreement with the exact evolution is obtained with the normal distribution.  

From this analysis, it is tempting to conclude that the Gaussian assumption is validated. However, the better agreement directly stems from the nature of the initial state. Indeed,        
coherent states have Gaussian (or Gaussian-like) shapes that give Gaussian-like distributions for observables. This conclusion can be drawn by analyzing the relative third and fourth centered moments generally called skewness and kurtosis of the initial state (see Appendix A for details). The quantum skewnesses and kurtoses can be easily obtained from Eq. (\ref{qusk}) as 
\begin{eqnarray}
\gamma^{(Q)}_x&=&\gamma^{(Q)}_y=-0.25\quad\gamma^{(Q)}_z=0.18,\nonumber\\
\beta^{(Q)}_x&=&\beta^{(Q)}_y=3.01\quad\;\;\,\beta^{(Q)}_z=2.98,
\end{eqnarray}
for the initial coherent state $|\theta_0=\pi/3,\phi_0=\pi/4\rangle$. The quantum skewnesses are close to zero and the quantum kurtoses are very close to 3 which is the kurtosis of normal distribution. For any initial Bloch coherent state $|\theta_0,\phi_0\rangle$, the quantum kurtoses are very close to the kurtosis of normal distribution, therefore normal distribution is the best choice for the initial distribution of collective observables when the initial state is a coherent state. This will be further illustrated below by a direct focus on the initial phase-space.

\subsubsection{SMF evolution starting from a correlated state}
The SMF approach has been originally formulated for initial states that are either pure independent particle states or 
statistical independent particle states. Here, we show that SMF can also be applied starting from a correlated state using exactly 
the same strategy. In that case, the initial phase-space is obtained by mapping the quantum average including correlation effect 
to classical average and then performing independent mean-field evolutions.

\begin{figure}[htb]
\includegraphics[width=7.5cm]{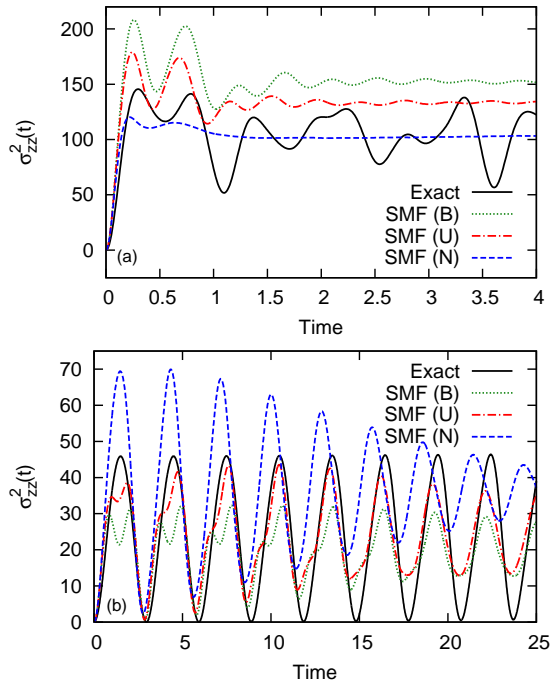}
\caption{(Color online) The variances of $J_z$ are plotted versus time for the initial non-coherent state $|j\;0\rangle$. B, U, and N signify bimodal, uniform, and normal distributions, respectively. The mean-field solution is constant and equal to 0, hence it is not shown. a) strong coupling ($\chi=5$) case. b) weak coupling ($\chi=0.5$) case.}
\label{fig:mixed1}
\end{figure}
Figure \ref{fig:mixed1} is the same with Figure \ref{fig:jz1} except that initial state is the correlated (non-coherent) state $|j\:0\rangle$. This state, being an eigenstate of the two-body operator $J^2$, is strongly entangled in terms of the coherent states. Using the closure relation Eq. (\ref{eq:closure}), this state can be written as
\begin{eqnarray}
|j\;0\rangle=\frac{2j+1}{4\pi}\int |\Omega\rangle \langle\Omega|j\;0\rangle d\Omega,
\end{eqnarray}
where the mixing coefficients $\langle\Omega|j\;0\rangle$ can be eventually obtained by inverting Eq. (\ref{Bloch}).

Starting from this correlated state, the mean-field solutions for quasispins are constant and equal to their initial values. Contrary to all previous applications where a simple Slater determinant was considered initially \cite{Lac12,Lac13,Lac14b}, it is observed that SMF approach for all three distribution functions fails in the weak coupling regime. In this regime, it is also observed that the dynamics is very sensitive to the assumption made for the distribution. Similar conclusions are drawn in the strong coupling regime. In particular, the Gaussian approximation appears to be less accurate. This can be phenomenologically understood from the quantum skewnesses and kurtoses which are obtained as
\begin{eqnarray}
\gamma^{(Q)}_x&=&\gamma^{(Q)}_y=0,\nonumber\\
\beta^{(Q)}_x&=&\beta^{(Q)}_y=\frac{3}{2}-\frac{4}{N(N+2)}\approx 1.4976,
\label{bimkur}
\end{eqnarray}
for the correlated initial state $|j\;0\rangle$. In particular, the kurtosis for the correlated state is far from the Gaussian limit.  

One of the motivations to introduce the bimodal distribution is that it is slightly more flexible than the others. In particular, not only the first two moments can be imposed to match the quantal ones, but also the third and fourth moments can be imposed simultaneously. In Fig. \ref{fig:mixed1}, the kurtosis of the bimodal distribution is chosen to be equal to the quantum kurtosis given in Eq. (\ref{bimkur}). It is seen that even though the bimodal distribution satisfies the quantum expectation values up to the fourth moment of quasispin operators $J_i$, the agreement between SMF (B) and the exact solution is poor for weak and strong coupling cases. This is due to the fact that, in general, non-coherent states lead to complicated distribution functions for observables that cannot be guessed.   

As a matter of fact, the disagreement observed in Fig. \ref{fig:mixed1} directly stems from the inadequacy of any assumed initial distribution to account for the complexity of the real distribution of the observables. This observation points out the necessity to improve the description of the initial phase-space without necessarily relying on a specific assumption for the initial distribution.  

\section{Initial Distribution of Observables from The Husimi Function}
In the present section, we show that standard techniques generally used to obtain the phase-space of a quantum objects can be advantageously combined with the SMF approach to significantly improve the description of complex fermionic many-body systems. The phase-space formulation of quantum many-body systems is generally done by using one of the Glauber-Sudarshan P, Wigner, or Husimi Q quasiprobability distribution functions (see for instance \cite{Gar00}). A stochastic extension of the mean-field description of bosonic systems, especially the Bose-Einstein condensate (BEC), is introduced by what is called truncated Wigner approximation (TWA) (see \cite{Prou} and references therein). In this approach, the mean-field evolution is given by the time-dependent Gross-Pitaevskii equation and stochastic sampling from the initial Wigner distribution function is performed \cite{Proukakis}. However, in general, Wigner as well as Glauber-Sudarshan functions can assume negative values while the Husimi function is strictly positive-definite hence it can be easily used for stochastic simulations. For that reason, we use the Husimi function in the present study. In Ref. \cite{Trimborn,Polk}, the initial Husimi distribution function for the Bose-Hubbard model is used for constructing the initial quantum fluctuations. The Gross-Pitaevskii equation is again evolved with an ensemble of initial values that are sampled from the Husimi function. Our approach, while having a similar strategy,  is different than the mentioned studies for two reasons: (i) The approach we proposed can be applied to fermionic as well as bosonic systems that can be expressed in terms of some set of collective one-body observables. (ii) We provide a method to obtain initial phase-space distribution of any set of one-body observables and evolve them with the corresponding mean-field equations. As a result, a connection with the previous discussion on the SMF approach with a predefined distribution function is clearly provided. 

\subsection{Transforming the distribution of phase-space variables to distributions of any observables}
We have seen in the previous discussions that the initial distribution of some desired observables (quasispins $j_x,j_y,j_z$ in here) cannot be guessed in general situations. Then, the only way to obtain the shapes of distributions of these observables seems to be through a quasiprobability distribution function of conjugate phase-space variables. Before we explain how the initial distribution of observables is obtained with the Husimi function, we provide some necessary background information. The Husimi function is defined as
\begin{eqnarray}
 Q_0(\Omega)=\langle \Omega |\rho_0|\Omega\rangle,
\label{husome}
\end{eqnarray} 
where $\rho_0=|\psi_0\rangle\langle\psi_0|$ is the initial density operator. The expectation value of an arbitrary operator $A$ in terms of the Husimi function is given by 
\begin{eqnarray}
 \langle\psi_0|A|\psi_0\rangle=\frac{2j+1}{4\pi}\int P_A(\Omega) Q_0(\Omega) d\Omega,
\label{expA}
\end{eqnarray}
where $P_A(\Omega)$ is the antinormally ordered Weyl symbol of operator $A$. It is defined by
\begin{eqnarray}
 A=\frac{2j+1}{4\pi}\int P_A(\Omega) |\Omega\rangle\langle\Omega |d\Omega.
\label{weyl}
\end{eqnarray}       
The Weyl symbols of quasispin operators read as
\begin{eqnarray}
 P_{J_i}(\Omega)&=&\frac{N+2}{N}\langle\Omega |J_i|\Omega\rangle,\label{correc1}\\
 P_{\widetilde{J_iJ_j}}(\Omega)&=&\frac{(N+2)(N+3)}{N^2}\langle\Omega| J_i|\Omega\rangle\langle\Omega| J_j|\Omega\rangle\nonumber\\
&&-\frac{N+2}{4}\delta_{ij},
\label{correc}
\end{eqnarray}  
where the expectation values are given by Eq. (\ref{firstmom}).

One of the main subtlety with the Husimi function is the fact that it is a completely positive-definite distribution although it is representing a quantum object. Indeed, the price to pay to use a positive distribution is the necessity to use a renormalized quantity $P_A$ instead of directly $\langle\Omega |A |\Omega\rangle$ in Eq. (\ref{expA}). As we will see below this subtle aspect has to be taken with care when coupling to SMF is performed.

The first advantage of the Husimi method is that it can directly give access to the probability distribution of any states without assuming a priori its shape. The initial distribution of the desired observables can be written in terms of the conjugate phase-space variables by using the properties of the Husimi function. In LMG model, the conjugate phase-space variables are $p=\cos\theta$ and $q=\phi$ which are the normalized relative population difference between the upper and lower single-particle states and the relative phase between the upper and lower states, respectively. The Husimi function, Eq. \eqref{husome} provides the distribution of these conjugate phase-space variables, that is the distribution of angles $\theta$ and $\phi$. These distributions are shown in the top panels of Figs. \ref{fig:hus1} and \ref{fig:hus2} respectively for the pure coherent state and correlated state considered previously.  

To make contact with the distribution of quasispins used in SMF and discussed previously, it is necessary to transform the two angles into the $j_i$ components. This transformation turns out to be less straightforward than it could be anticipated due to the use of the Weyl symbols in quantum averages. The following strategy has been used here. First, an ensemble of stochastic angles $\{\theta^\lambda,\phi^\lambda\}$ that are sampled according to the Husimi function can be formed by using the rejection method. At this point, a method that transforms the ensemble of stochastic angles $\{\Omega^\lambda=(\theta^\lambda,\phi^\lambda)\}$ to an ensemble of stochastic quasispins $\{j_x^\lambda(0),j_y^\lambda(0),j_z^\lambda(0)\}$ is required. 
The direct approach of defining the quasispins as $j_i^\lambda(0)=P_{J_i}(\Omega^\lambda)$ where the Weyl symbol is obtained by substituting the stochastic angles into Eq. (\ref{correc1}) will give the correct quantum means when averaging over the ensemble is performed but the variances will be different than the quantum ones, since $P_{J_i}P_{J_j}\neq P_{\widetilde{J_iJ_j}}$. This problem can be resolved by defining the quasispins in an effective way. Let us write the quasispin components as an average and fluctuating part,
\begin{eqnarray}
j^\lambda_i(0) & = & \left[j_i^\lambda(0)\right]_H + \delta j_i^\lambda(0),
\end{eqnarray}
where $[...]_H$ means averaging over the Husimi distribution $Q_0$. Guided by the SMF approach, the properties 
of the average and fluctuating part in the above equation can directly be deduced imposing that they match the 
quantum ones associated to the considered state. The first moment directly gives
\begin{eqnarray}
\left[j_i^\lambda(0)\right]_H&=&\langle\psi_0|J_i|\psi_0\rangle = \left[P_{J_i}(\Omega^\lambda)\right]_H,
\end{eqnarray}
while the second moment imposes the constraint
\begin{eqnarray}
\left[\left(\delta j_i^\lambda(0)\right)^2\right]_H&=&\langle\psi_0|J_i^2|\psi_0\rangle-\langle\psi_0|J_i|\psi_0\rangle^2 \nonumber \\
&=&\left[P_{J_i^2}(\Omega^\lambda)\right]_H-\left[P_{J_i}(\Omega^\lambda)\right]_H^2, 
\end{eqnarray}
where Eq. \eqref{expA} has been used.  

Assuming that the fluctuations in the new effective quasispins are directly proportional to the fluctuations in the Weyl symbols, finally gives
\begin{eqnarray}
\delta  j_i^\lambda(0) &=& \sqrt{\frac{\left[P_{J_i^2}(\Omega^\lambda)\right]_H-\left[P_{J_i}(\Omega^\lambda)\right]_H^2}{\left[\left(\delta P_{J_i}(\Omega^\lambda)\right)^2\right]_H}}\;\delta P_{J_i}(\Omega^\lambda), 
\label{husqs}
\end{eqnarray}
where $\delta P_{J_i}(\Omega^\lambda)= P_{J_i}(\Omega^\lambda)- \left[P_{J_i}(\Omega^\lambda)\right]_H$. In summary, the quasispin phase-space associated to a given initial state can be obtained using the numerical scheme:
\begin{eqnarray}
 \{ (\theta^\lambda,\phi^\lambda)\}~~\longrightarrow ~~  \{ P_{J_i}(\Omega^\lambda),P_{J_i^2}(\Omega^\lambda)\}     ~~\longrightarrow ~~\{ j_i^\lambda(0) \},  \nonumber 
\end{eqnarray}
where the angles are sampled according to the Husimi function, Eq. \eqref{husome}. Then, the associated set of Weyl symbols are computed from Eqs. \eqref{correc1} and \eqref{correc}. Finally, the set of quasispins are deduced from Eq. (\ref{husqs}). The great advantage of the present method is that it provides automatically the distribution of quasispins with first and second moments that match the exact quantum ones, hence removing the need for guessing the shape of the initial probability distribution.  

\subsection{Initial phase-space associated to coherent and correlated states}
We illustrate here the strength of the approach for the two initial states considered previously. The Husimi functions for an initial coherent state 
$|\theta_0,\phi_0\rangle$ and for an initial correlated state $|j\: m\rangle$ are given by 
\begin{eqnarray}
 Q_0(\theta,\phi)\!&=&\!|\langle\theta,\phi|\theta_0,\phi_0\rangle|^2\nonumber\\
          &=&\!\!\!\left[\frac{1+\cos\theta\cos\theta_0+\sin\theta\sin\theta_0\cos(\phi-\phi_0)}{2}\right]^{2j}, 
\label{hus1}
\end{eqnarray}  
and
\begin{eqnarray}
 Q_0(\theta,\phi)&=&|\langle\theta,\phi|j\: m\rangle|^2\nonumber\\
          &=&{2j \choose j-m}\left(\frac{1+\cos\theta}{2}\right)^{j+m}\left(\frac{1-\cos\theta}{2}\right)^{j-m},
\label{hus2}
\end{eqnarray} 
respectively. 
\begin{figure}[htb]
\includegraphics[width=7.5cm]{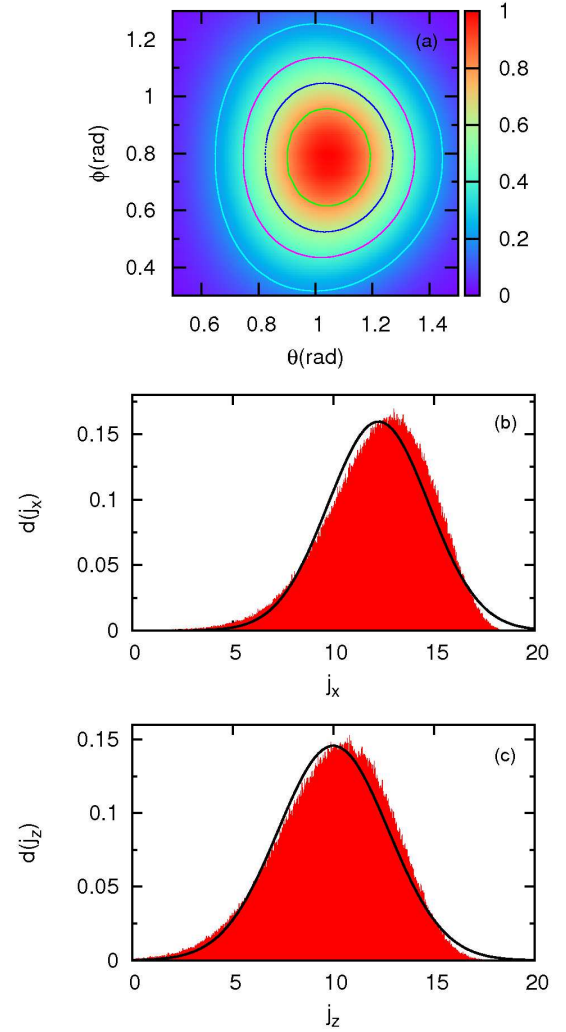}
\caption{(Color online) a) The Husimi function Eq. (\ref{hus1}) for the initial coherent state $|\pi/3,\pi/4\rangle$ is shown. In this plot, Husimi function is normalized so that its maximum value is 1. Figures b) and c) show distributions of $j_x$ and $j_z$, respectively. The filled areas indicate distributions that are obtained according to the Husimi function whereas black lines indicate the normal distributions.}
\label{fig:hus1}
\end{figure}
\begin{figure}[htb]
\includegraphics[width=7.5cm]{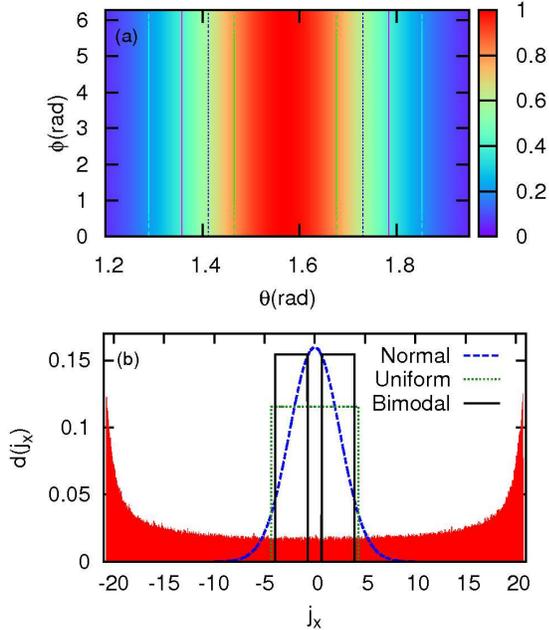}
\caption{(Color online) a) The Husimi function given by Eq. (\ref{hus2}) is shown. In this plot, Husimi function is normalized so that its maximum value is 1. b) The distribution of $j_x$ is shown. The filled area indicates the distribution that is obtained according to the Husimi function. The three distribution functions are also indicated. The kurtosis for the bimodal distribution is the value given by Eq. (\ref{bimkur}).}
\label{fig:hus2}
\end{figure}

Figure \ref{fig:hus1}a shows the shape of the Husimi function for the initial coherent state $|\pi/3,\pi/4\rangle$. It is seen that the shape is close to that of a Gaussian. Figures \ref{fig:hus1}b and \ref{fig:hus1}c indicate the distributions of quasispins $j_x$ and $j_z$, respectively. The distribution of $j_y$ is the same with that of $j_x$. The quasispin distributions obtained from the Husimi function are close to Gaussian shapes. This confirms the skewness and kurtosis analysis made previously and also explains why in this case, the Gaussian assumption made at initial stage of SMF was more predictive than the other distributions. In Fig. \ref{fig:hus2}a, the shape of the Husimi function for the initial correlated state $|j\: 0\rangle$ is shown. The Husimi function is independent of $\phi$, hence $\phi$ is uniformly distributed. Figure \ref{fig:hus2}b shows the distribution of $j_x$. The distribution of $j_y$ is the same with that of $j_x$. The $z$ component is not distributed since its quantum variance is zero, hence it has a single initial value. Note that even though the first four moments of quasispins obtained from the bimodal (B) distribution (black line) and Husimi function (red filled area) are the same, the shapes of these two distributions are completely different. From Fig. \ref{fig:hus2}, it is clear that (i) none of the assumed initial distributions is close to the real phase-space, (ii) this phase space could have hardly been guessed without the present analysis.  

\subsection{Combining the Husimi phase-space sampling and SMF approach}
\begin{figure}[htb]
\includegraphics[width=7.5cm]{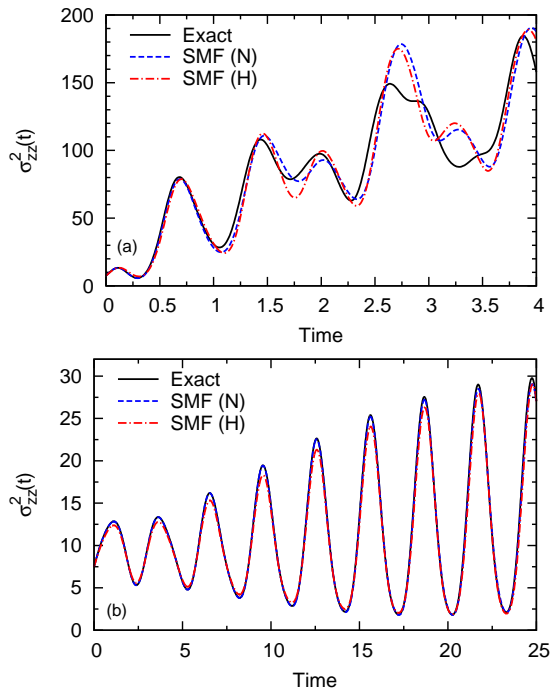}
\caption{(Color online) The variances of $J_z$ are plotted versus time for the initial coherent state $|\theta_0=\pi/3,\phi_0=\pi/4\rangle$. The exact variances are compared with the approximate ones obtained with SMF (N) and SMF (H) methods. a) strong coupling ($\chi=5$) case. b) weak coupling ($\chi=0.5$) case.}
\label{fig:husdyn1}
\end{figure}

\begin{figure}[htb]
\includegraphics[width=7.5cm]{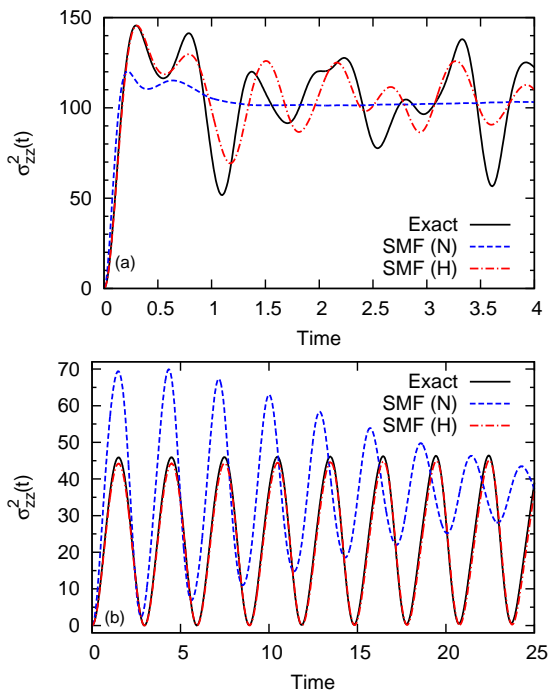}
\caption{(Color online) The variances of $J_z$ are plotted versus time for the initial non-coherent state $|j\;0\rangle$. The mean-field solution is constant and equal to 0, hence it is not shown. a) strong coupling ($\chi=5$) case. b) weak coupling ($\chi=0.5$) case.}
\label{fig:husdyn2}
\end{figure}

We now return to one of the main motivation of the present work, and show that a realistic description of the initial phase-space can extend the domain of validity of SMF and enable to describe the dynamics of an initially correlated state. The sampling method of initial quasispins based on the Husimi function is now used to obtain the set of initial configuration for the mean-field dynamics. In Fig. \ref{fig:husdyn1}, the exact variances of $J_z$ as well as the variances computed by using SMF approach with normal (N) distribution and with distribution provided by the Husimi (H) function are shown. It is seen that both SMF (N) and SMF (H) provide good agreement with the exact variances at weak and strong coupling regimes for the initial coherent state $|\pi/3,\pi/4\rangle$. Systematic analysis was performed and it was observed that, for any initial coherent state, observables computed by using SMF (N) and SMF (H) approaches provide good agreement with the exact observables. Figure \ref{fig:husdyn2} shows the time evolution of the variances of $J_z$ for the initial non-coherent state $|j\;0\rangle$. In the weak coupling regime, it is observed that SMF approach with distribution obtained by the Husimi (H) function provides a very good approximation to the exact evolution and therefore considerable improvement over the results of Fig. \ref{fig:mixed1} where the distributions have been predefined is achieved. Therefore, we see that a more realistic description of the initial phase-space has a direct impact on the predictive power of SMF approach. We performed systematic studies with different non-coherent states and observed that SMF (H) always gives a better description of the time evolution of observables than SMF (N) approach. 

It is interesting to note from Fig. \ref{fig:husdyn2} as well as Fig. \ref{fig:mixed1} that the SMF theory applied with Gaussian approximation leads 
to an overdamping of the fluctuation compared to the exact case in the strong coupling regime. This has been misleadingly interpreted in Refs. \cite{Lac12,Lac13,Lac14b} as a generic defect of SMF. These two figures clearly demonstrate that this overdamping can disappear simply by changing the 
initial distribution.

\subsection{SMF with an initial mixed density}
\begin{figure}[htb]
\includegraphics[width=7.5cm]{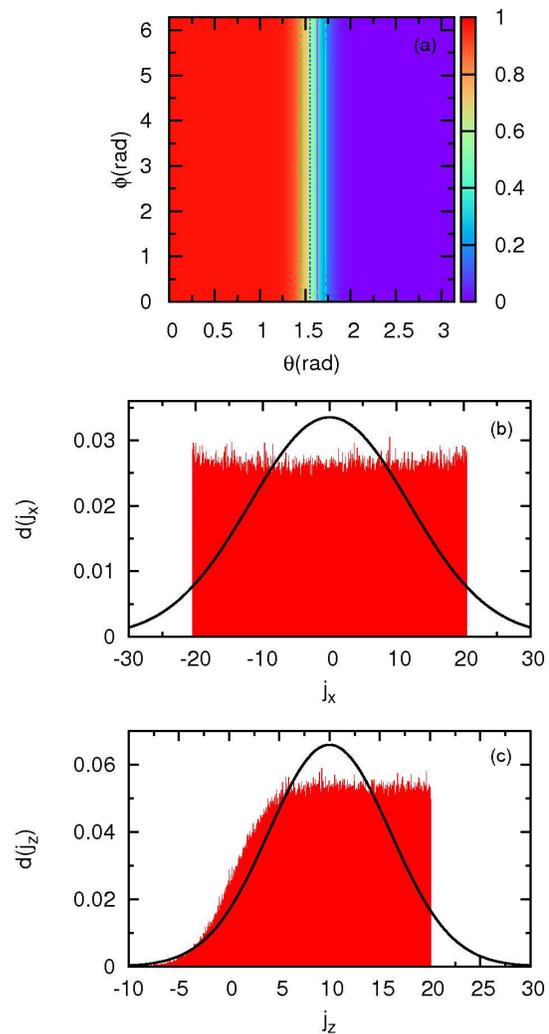}
\caption{(Color online) a) The Husimi function given by Eq. (\ref{hus3}) is shown. In this plot, Husimi function is normalized so that its maximum value is 1. Figures b) and c) show distributions of $j_x$ and $j_z$, respectively. The distribution of $j_y$ is the same with that of $j_x$. The filled area indicates the distribution that is obtained according to the Husimi function. The normal distributions are also indicated.}
\label{fig:mix1}
\end{figure}

\begin{figure}[htb]
\includegraphics[width=7.5cm]{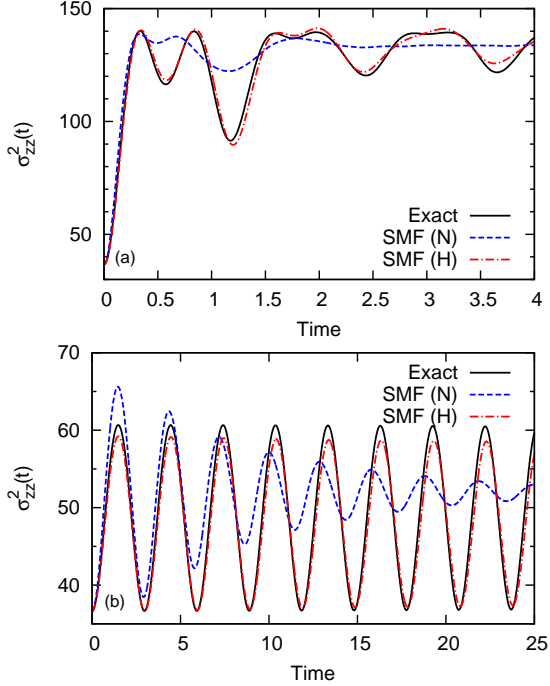}
\caption{(Color online) The variances of $J_z$ are plotted versus time for the initial mixed state Eq.(\ref{mixedst}). The mean-field solution is constant and equal to 0, hence it is not shown. a) strong coupling ($\chi=5$) case. b) weak coupling ($\chi=0.5$) case.}
\label{fig:mix2}
\end{figure}

As an additional illustration of the importance of initial sampling, we consider a situation where the initial density cannot be considered as
a pure state density. We consider an initial density written as a mixing of several states,
\begin{equation}
 \rho_0=\frac{1}{j+1}\sum_{m=0}^j|j\;m\rangle\langle j\;m|.
\label{mixedst}
\end{equation}
The corresponding Husimi function is obtained as
\begin{eqnarray}
 Q_0(\theta,\phi)=\frac{1}{j+1}\sum_{m=0}^j&&{2j \choose j-m}\left(\frac{1+\cos\theta}{2}\right)^{j+m}\nonumber\\
                 &&\times\left(\frac{1-\cos\theta}{2}\right)^{j-m}
\label{hus3}
\end{eqnarray}
and is shown in Fig. \ref{fig:mix1}a. As we see in this figure, the $\phi$ component is uniformly distributed while the $\theta$ component has a Fermi shape. Therefore none of the two angles resembles a Gaussian distribution. In Fig. \ref{fig:mix2}, the SMF evolution starting from the mixed density using either a Gaussian assumption or the sampling based on the Husimi technique is compared with the exact solution. While the former again fails to provide the right evolution, the latter gives very good results both in the weak and strong coupling regime.  


\section{Conclusions}
In the present work, we investigate the validity and sensitivity of the Gaussian approximation made in the recently proposed stochastic mean-field 
approach \cite{Ayi08}. This approach, where the initial phase-space of a selected collective variable set is sampled to get a set of initial conditions followed by standard mean-field dynamics, has been shown to tremendously improve the independent particle picture by including the effect of quantal fluctuations beyond mean-field. Using the Lipkin-Meshkov-Glick model as an illustration, we show here that indeed, the Gaussian assumption is a good approximation if the initial state is not correlated as was always the case in all applications up to now \cite{Lac12, Lac13, Lac14b}. However, when the initial state is correlated and corresponds to a mixing of Slater determinants, SMF dynamics with the Gaussian approximation for the initial distributions can fail even in the weak coupling regime. We explicitly demonstrate that the failure comes from the complexity of the initial phase-space that cannot be assumed to be Gaussian anymore. This finding motivates the search for more accurate methods to describe initially correlated states. It is proposed here to combine the SMF approach with more elaborate technique to obtain the initial phase-space of a quantum state. An illustration is given here using the Husimi distribution that leads to a more realistic description of the initial conditions. A methodology is proposed to use this technique within the SMF. We show that a more realistic description of the initial phase-space greatly improves the description of correlated systems. Application to the LMG model gives perfect results in the weak coupling regime while in the strong coupling regime, the time over which the SMF dynamics is in agreement with the exact solution is increased compared to the simple Gaussian sampling assumption. The present study demonstrates that the predictive power of the SMF approach can further be improved if a special attention is made on the initial phase-space and approximation beyond the Gaussian sampling is made. This might in particular be crucial, when correlated initial quantum states are considered. Here, we show how the Husimi technique can be used to get a more realistic phase-space. This technique can be very helpful in situations where the coherent state algebra is well under control. Such Husimi approach has been recently applied in Ref. \cite{Loe12}. Note that other techniques like Wigner transform that allow to make connection between quantum and classical mechanics can be used. Such Wigner functions have been, for instance, used in the nuclear context for a different purpose \cite{Loe11,Yil14}. 
              
\begin{acknowledgments}
B.Y. gratefully acknowledges IN2P3-CNRS, Universit\'e Paris-Sud for warm hospitality extended to him during his visit.
\end{acknowledgments}

\appendix
\section{}
The normal distribution function with means $\mu_i$ and (co)variances $\sigma_{ij}^2$, is given by
\begin{eqnarray}
D_N(j_x,j_y,j_z)&=&\frac{1}{\sqrt{(2\pi)^3|\bm{\Sigma}|}}\nonumber\\
&&\times\exp\left[-\frac{1}{2}(\bm{j}-\bm{\mu})^T\bm{\Sigma}^{-1}(\bm{j}-\bm{\mu})\right],
\end{eqnarray}
where $\bm{j}=(j_x,j_y,j_z)$ and $\bm{\mu}=(\mu_x,\mu_y,\mu_z)$ are vectors and $\bm{\Sigma}$ is the variance-covariance matrix given by
\begin{equation}
 \bm{\Sigma}=\left(\begin{array}{ccc} 
\sigma_{xx}^2 & \sigma_{xy}^2 & \sigma_{xz}^2 \\
\sigma_{xy}^2 & \sigma_{yy}^2 & \sigma_{yz}^2 \\
\sigma_{xz}^2 & \sigma_{yz}^2 & \sigma_{zz}^2 \\
\end{array}\right).
\end{equation}
$|\bm{\Sigma}|$ is the determinant of the matrix $\bm{\Sigma}$. The uncorrelated uniform distribution function is given by
\begin{eqnarray}
D_U(j_x,j_y,j_z)=\prod_{i}d_U(j_i),
\end{eqnarray}
where  
\begin{eqnarray}
d_U(j_i)=\left\{\begin{array}{cc}
                 \frac{1}{2\sqrt{3\sigma^2_{ii}}}, & \text{ if }\; |j_i-\mu_i|<\sqrt{3\sigma^2_{ii}} \\
                 0, & \text{  otherwise.} \\
                \end{array}\right.
\end{eqnarray}
We also arbitrarily define a bimodal distribution that is given by
\begin{eqnarray}
D_B(j_x,j_y,j_z)=\prod_{i}d_B(j_i),
\end{eqnarray}
with  
\begin{eqnarray}
d_B(j_i)=\left\{\begin{array}{ccc}
                 \frac{1}{2(b_i-a_i)}, & \text{if\; } a_i<j_i-\mu_i<b_i \\
                 \frac{1}{2(b_i-a_i)}, & \text{if } -b_i<j_i-\mu_i<-a_i \\
                 0, & \text{otherwise}. \\
                \end{array}\right.
\end{eqnarray}
For this distribution, the variances can be easily obtained in terms of $a_i$ and $b_i$ as
\begin{equation}
 \sigma^2_{ii}=\frac{a_i^2+a_ib_i+b_i^2}{3}.
\label{bivar}
\end{equation}
 
For all three distributions, if the covariances $\sigma^2_{ij}$ are non-zero, correlated random variables can be produced from uncorrelated ones as explained in Appendix B. 

Each of the normal and uniform distribution functions depends on two sets of parameters $\mu_i$ and $\sigma_{ij}^2$ which can be fixed by choosing them to be equal to the quantum means and (co)variances, that is $\mu_i=\langle\psi_0|J_i|\psi_0\rangle$ and $\sigma^2_{ij}=\langle\psi_0|\widetilde{J_iJ_j}|\psi_0\rangle-\langle\psi_0|J_i|\psi_0\rangle\langle\psi_0| J_j|\psi_0\rangle$. On the other hand, the bimodal distribution depends on three sets of parameters $\mu_i$, $a_i$, and $b_i$. Therefore, next to the quantum means and (co)variances, in principle, the expectation values of higher moments of observables can also be reproduced. 

The third and fourth moments of a distribution are generally expressed in terms of the skewness $\gamma_i=[(j_i-\mu_i)^3]_{D}/(\sigma^2_{ii})^{3/2}$ and the kurtosis $\beta_i=[(j_i-\mu_i)^4]_{D}/(\sigma^2_{ii})^{2}$. The skewnesses and kurtoses of the three distributions are given by
\begin{eqnarray}
\begin{array}{ccc}
\gamma^{(N)}=0 & \beta^{(N)}=3 & \text{(normal),}\\
\gamma^{(U)}=0 & \beta^{(U)}=1.8 & \text{(uniform),}\\
\gamma^{(B)}=0 & 1<\beta^{(B)}<1.8 & \text{(bimodal).}
\end{array}
\label{gambet}
\end{eqnarray}
The kurtosis of the bimodal distribution depends on the parameters $a_i$, and $b_i$. It is obtained as 
\begin{equation}
 \beta^{(B)}_i=\frac{9}{5}\frac{a_i^4+a_i^3b_i+a_i^2b_i^2+a_ib_i^3+b_i^4}{(a_i^2+a_ib_i+b_i^2)^2}.
\label{bikur}
\end{equation}
For any given variance and kurtosis which is within the interval provided in Eq. (\ref{gambet}), the parameters $a_i$ and $b_i$ can be calculated by solving Eqs. (\ref{bivar}) and (\ref{bikur}). Hence, the bimodal distribution introduces some freedom for the choice of kurtosis.
The quantum skewness and kurtosis of the observables $J_i$ can be similarly defined for the initial state $|\psi_0\rangle$ as 
\begin{eqnarray}
\gamma^{(Q)}_i&=&\frac{\langle\psi_0|(\Delta J_i)^3|\psi_0\rangle}{\langle\psi_0|(\Delta J_i)^2|\psi_0\rangle^{3/2}},\nonumber\\
\beta^{(Q)}_i&=&\frac{\langle\psi_0|(\Delta J_i)^4|\psi_0\rangle}{\langle\psi_0|(\Delta J_i)^2|\psi_0\rangle^{2}}, 
\label{qusk}
\end{eqnarray}
where $\Delta J_i=J_i-\langle\psi_0|J_i|\psi_0\rangle$. The quantum skewnesses and kurtosis can be compared to the ones for the normal, uniform, and bimodal distributions. This will allow us to understand which distribution is closer to the actual distribution for a given initial state. 

\section{}
Correlated random variables can be produced by using uncorrelated random variables. Let $Y$ represent the vector of the desired correlated random variables with means represented by the vector $\mu$ and (co)variances represented by the matrix $\Sigma$. If $Z$ is a vector of uncorrelated random variables with zero mean and unit variance, it is possible to obtain $Y$ from $Z$ by the equation
\begin{align}
Y&=\mu + CZ\nonumber\\
\left(\begin{array}{c} Y_1 \\ Y_2 \\ \vdots \\ Y_k \end{array}\right)&=\left(\begin{array}{c} \mu_1 \\ \mu_2 \\ \vdots \\ \mu_k \end{array}\right)\! +\! \left(\begin{array}{cccc} 
C_{11} & C_{12} & \cdots & C_{1k} \\
C_{21} & C_{22} & \cdots & C_{2k} \\
\vdots & \vdots & \ddots & \vdots \\
C_{k1} & C_{k2} & \cdots & C_{kk} \\
\end{array}\right)\!\!\!  
\left(\begin{array}{c} Z_1 \\ Z_2 \\ \vdots \\ Z_k \end{array}\right) 
\end{align}
for $k$ random variables. It is obvious that $\left[Y\right]_D=\mu$ since $\left[Z\right]_D$ is zero. The variance of the vector $Y$ reads as
\begin{eqnarray}
\left[YY^T\right]_D&=&\mu\mu^T+C\left[ZZ^T\right]_DC^T\nonumber\\
               &=&\mu\mu^T+\Sigma, 
\end{eqnarray}
where $\left[ZZ^T\right]_D$ is unit matrix and 
\begin{eqnarray}\label{Chol}
\Sigma&=&CC^T\nonumber\\
 &=&\left(\begin{array}{cccc} 
\sigma^2_{11} & \sigma^2_{12} & \cdots & \sigma^2_{1k} \\
\sigma^2_{21} & \sigma^2_{22} & \cdots & \sigma^2_{2k} \\
\vdots & \vdots & \ddots & \vdots \\
\sigma^2_{k1} & \sigma^2_{k2} & \cdots & \sigma^2_{kk} \\
\end{array}\right)
\end{eqnarray}
is a given variance-covariance matrix. The matrix $C$ in Eq. (\ref{Chol}) can be obtained from the variance-covariance matrix by using Cholesky decomposition or eigendecomposition.

\end{document}